# Broadband optical time-domain reflectometry for security analysis of quantum key distribution

Klim D. Bondar[*,1,2], Ivan S. Sushchev[1,2], Daniil D. Bulavkin[1], Kirill E. Bugai[1], Anna S. Sidelnikova[1], Dmitriy M. Melkonian[1], Veronika M. Vakhrusheva[1,2], and Dmitriy A. Dvoretskiy[1]

[1]*SFB Laboratory, LLC, 127273 Moscow, Russia*
[2]*Quantum Technology Centre and Faculty of Physics, Lomonosov Moscow State University, 119991 Moscow, Russia*
(Dated: 2025-07-24)

In this paper, we present a method for security analysis against the Trojan-horse attacks (THA) launched to a practical fiber-based quantum key distribution (QKD) system across a wide spectral range. To achieve this, we utilize optical time-domain reflectometry (OTDR) for the spectral reflectance analysis in the near-infrared range $\lambda = 1100 - 1800$ nm with centimeter-level resolution and with a noise floor down to -80 dB. Finally, the total theoretical-security analysis against the THA side channel considering the spectral reflectance and transmittance data over this spectral range is conducted. To the best of our knowledge, our OTDR setup and the corresponding results are the first of their kind in a wide spectral range.

## I. INTRODUCTION

Currently, quantum cryptography stands as one of the fastest-growing applications of quantum information theory, leading to the emergence of real quantum key distribution (QKD) systems. The majority of QKD protocols provide unconditional security relying on the laws of quantum physics [1]-[3]. However, establishing a theoretical framework that considers all physical effects within QKD systems, including imperfections in the optical components, is a challenging task. As a result, various types of attacks exploiting implementation loopholes have emerged as a concern [4]-[7].

One such vulnerability is the non-zero reflection from the interfaces inside a QKD system. Exploiting this, an eavesdropper (Eve) could attempt extracting secret key bits by launching bright light pulses into the setups of the legitimate users (Alice or Bob). She could try to retrieve the prepared legitimate quantum state (attacking Alice's setup) as well as the detection basis (attacking Bob's setup) from the reflected and modulated photons. This type of attack is known as the Trojan-horse attack (THA)[8]-[19].

The primary methods to mitigate THA involve passive and active defense components, such as optical circulators and isolators, which transverse light in specific directions; optical attenuators that limit the power of passing radiation; bandpass filters installed to suppress the non-legitimate light by limiting the radiation spectrum; watchdog detectors capable of identifying non-legitimate pulses. Nevertheless, their effectiveness depends on the spectrum of incoming light, which may be arbitrarily selected by Eve. Thus, Eve can launch an attack within a wavelength range, wherein she remains invisible to legitimate users notwithstanding watchdog detectors as well as the passive components are in use. As a result, considering the efficiency of these passive methods across a wide spectral range is essential. Such measurements have been conducted successfully, yielding valuable information on the transmission of different combinations of passive optical defense components within a wide spectral range [10], [15]-[19]. However, to mitigate THA, information about the reflection from the optical components within the optical setup is required. In the absence of the spectral reflection values, one must suppose the non-physical total spectral reflection from the QKD setup while conducting security analysis.

In our research, we employ a well-established optical time-domain reflectometry (OTDR) method, commonly used in conventional fiber optics for return losses measurements [20]. In most cases typical reflections from QKD systems apparatuses lie in the $-30$ dB $\div$ $-60$ dB range and thus are substantially faint, so the significant sensitivity of the detection is required. An appropriate dynamic range for these measurements can be obtained by using single-photon avalanche photodiodes operated in Geiger mode (the so-called $\nu -$ OTDR [21]-[22]). Reflectometry setups that operate at specific wavelengths are widely used in conventional fiber optics. Reflectometry pictures of the QKD setups on the separate wavelengths have been presented in Refs. [11]-[12], [14]. Nevertheless, reflectometry measurements across the wide spectral range have never been reported. In the context of the broadband THA security analysis, the reflectance estimation was either incomplete [11]-[12], [14], [16] or required a non-physical assumption on total reflection of Eve radiation from QKD setup parts [19][19], [25].

---

[*]bondar.kd19@physics.msu.ru

Note that there are methods of optical reflectometry in frequency domain (OFDR), wherein heterodyne detection is combined with linear sweep of the laser optical frequency, producing reflectometry traces in frequency domain with peaks corresponding to beat frequencies [22]. Then one can obtain classical reflectometry traces in time domain by Fourier-transform analysis. Owing to heterodyne detection technique, the dynamic range of today's commercially available OFDR setups is down to -100 dB with a micrometer-level spatial resolution, which has never been performed by conventional OTDR techniques. Nevertheless, OFDR appears to be more complicated procedure compared to OTDR: the nonlinearity of realistic frequency sweeping, the finite coherence time, intensity fluctuations and phase noise factors of the laser source, as well as polarization misalignments between local oscillator signal and reflected/backscattered signal, should be considered and optimized [24]. Nevertheless, applying to QKD-security analysis against THA one needs only to measure typical reflections ranged from $-30$ dB to $-60$ dB and distinguish them at centimeter-level distances between different fiber components. The suitability of $\nu$−OTDR technique for such experimental analysis have been shown previously in Refs. [12], [14], [19].

We have successfully developed an OTDR setup based on a picosecond supercontinuum laser source and an avalanche InGaAs photodiode. This allowed us to conduct OTDR measurements on a QKD system along a wide spectral range for the first time, to the best of our knowledge. The rest of this paper is organized as follows. Sec. II represents an overview of the Trojan-Horse attack and experimental methods for security evaluation against it. There, we also present our broadband OTDR setup and describe its operation. In Sec. III the broadband OTDR traces over a practical fiber-based QKD system are presented across the wavelength range $\lambda = 1100 - 1800$ nm. Here, we also experimentally show that the largest reflections ($\approx -50$ dB) in our fiber-based QKD system under test occur from the optical connectors. Then we additionally assess return losses of fiber PC-type connectors by conducting OTDR on them individually and compare the corresponding OTDR-picture with the theoretical connectors return losses curve across the aforementioned spectral range. In Sec. IV, we finally conduct the theoretical-security analysis of our QKD system and show that it is sustainable to a broadband THA owing to defense components in use. In Sec. V we conclude our paper and highlight the key points in assessing the robustness of an arbitrary QKD system against broadband THA.

## II. METHODS FOR THA SECURITY ANALYSIS

### A. Information leakage to an eavesdropper

By launching the THA to an arbitrary QKD system, Eve sends bright light pulses to Alice's or Bob's optical setup (see FIG. 1). Her possibility to reveal secret key bits depends on the distinguishability of the quantum states for different bits. For instance, in the case of the phase-coded BB84 protocol [1]-[2] the information leakage to Eve via the THA side channel can be estimated by using the lower bound of the square root of the Uhlmann fidelity $\eta$ between Trojan quantum states returned to Eve with phases displaced on $\pi$. It can be explicitly shown [see Appendix] that the lower square root fidelity bound for phase-coded states is the same for pure and mixed states, and thus we can consider THA states to be pure. Furthermore, we should assume that all errors during the QKD session are originated by Eve, i.e. her collective attack. It has been shown that collective attacks are optimal for the protocols with independent and permutable channel uses [5], which is our case. In case of optimal symmetric collective attack [6], we should account for a factor of the fidelity square root $\varepsilon$ between two types of perturbed states $|\Phi_0\rangle, |\Phi_1\rangle$ and $|\Theta_0\rangle, |\Theta_1\rangle$ of Eve's auxiliary quantum subsystem [8]. This leakage is limited by the upper bound of Holevo quantity $\bar{\chi}_{\text{Eve}}$ [19] as follows:

$$\eta \approx 1 - 2\mu_{\text{Eve}}, \quad (1)$$

$$\varepsilon = \langle\Phi_0|\Phi_1\rangle = \langle\Theta_0|\Theta_1\rangle = 1 - 2Q, \quad (2)$$

$$\bar{\chi}_{\text{Eve}} = h\left(\frac{1-\eta\varepsilon}{2}\right) \approx h\left(\frac{1-(1-2Q)\cdot(1-2\mu_{\text{Eve}})}{2}\right), \quad (3)$$

wherein $\mu_{\text{Eve}}$ is the mean photon number in a pulse returned to Eve, $Q$ is a quantum bit error rate (QBER) produced by Eve and $h(x) = -x \cdot \log_2 x - (1-x) \cdot \log_2(1-x)$ is the binary entropy function.

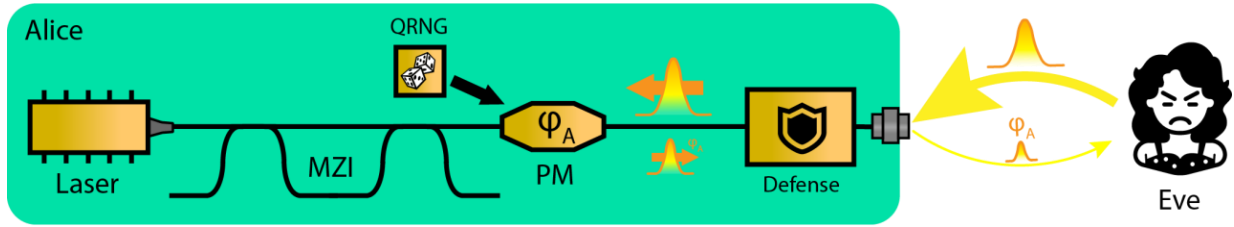

FIG. 1. Simplified demonstration of the THA on Alice's setup. *Laser*, Alice's laser; *PM*, Alice's phase modulator; *MZI*, Mach-Zehnder interferometer; *QRNG*, quantum random number generator; *Defense*, passive optical defense components

Therefore, the analysis of information leakage to Eve via the THA side channel comes down to the estimation of the value $\mu_{Eve}$, that is related to the average power of radiation reflected to Eve $P_{\text{refl}}$:

$$\mu_{\text{Eve}}(\lambda) = \frac{P_{\text{Eve}}(\lambda)[W] \cdot \lambda}{f_{\text{Eve}} h c}, \quad (4)$$

where $h \approx 6.63 \cdot 10^{-34} \, J \cdot Hz^{-1}$ is the Planck constant, $\lambda$ is the wavelength of reflected radiation, $c \approx 3 \cdot 10^8 \, m/s$ is the velocity of light, $f_{\text{Eve}}$ is Eve's pulses repetition frequency. The $P_{\text{Eve}}$ value is in turn related to the maximum acceptable input power of Eve's radiation $P_{\text{max}}$:

$$P_{\text{Eve}}[\text{dBm}] = P_{\text{max}}[\text{dBm}] + T[\text{dB}] + R[\text{dB}]. \quad (5)$$

The technique for the spectral transmittance estimation has already been presented [19]. The input power of Eve's radiation should also be considered. The power cannot be arbitrarily large – its upper limit is determined by the laser-induced damage threshold [14], otherwise, the QKD operation breaks, and THA fails. If watchdog detectors are used, the upper limit of Eve's input radiation corresponds to the detector sensitivity threshold, which can be experimentally assessed along the spectral range of the inducing radiation. Finally, the spectral reflectance can be assessed by the broadband OTDR technique presented in this paper.

**B. OTDR measurements setup**

The essential part to account for the reflection values from all basic components of the device under test with sufficient accuracy is achieving the maximum spatial resolution and dynamic range of the measurements while conducting $\nu - $ OTDR. Here, we present our experimental setup (FIG. 2 (a)) which satisfies both requirements of maximum achievable spatial resolution and dynamic range of measurements. A supercontinuum laser source (SCL) with several hundred picoseconds pulse duration provides the centimeter-level resolution (FIG. 2 (b)). The spatial resolution data was obtained by the reflection peak half-width from the measured OTDR-picture. In turn, the single-photon detector based on the InGaAs photodiode (SPAD) provides sufficient sensitivity and low noise over the spectral range $\lambda = 1100 — 1800$ nm. Nevertheless, the quantum efficiency of InGaAs is nonzero but less than 1% at boundary wavelengths points (i.e. 1100 nm, 1800 nm) for the spectral region studied. Thus, reflection peaks are highly affected by noise (see FIG. 2 (c)), and the spatial resolution values appear to be approximate at wavelengths points close to boundary ones. At another extreme, the current reflection values at boundary wavelengths points are about 10 dB higher than -80 dB noise floor and thus are not optimal for the THA launching. One way to reduce this effect is increasing the measurements duration, if the current wavelength appears to be optimal in context of the THA.

Light pulses from the SCL pass through the tunable acousto-optical filter (TF) installed to set the measurements wavelength. This way, by gradually tuning the wavelength of the TF, the scan over studied spectral range was achieved. Our filter operates within 1100-2000 nm wavelength range with 6~12 nm bandwidth, and for our measurements we applied 25 nm wavelength scanning step. Then, the spectrally selected radiation is attenuated by the variable optical attenuator (VOA) to decrease the optical power to be properly detected on SPAD (see sec. III (A) for the conditions of the appropriate detection). After passing through the $1 \rightarrow 2$ circulator direction, the radiation reflects from the optical components located inside the device under test (DUT) to be detected by the SPAD after passing through the $2 \rightarrow 3$ circulator direction. The time-controlling electronics perform the sampling in the time domain during the period of laser pulses repetition. The photon counts are accumulated within the specific time bins with 150 ps duration.

The reflection values from the DUT components can be determined by the assessment of the input power $P_{\text{in}}$ and the reflected power $P_{\text{out}}$ as follows:

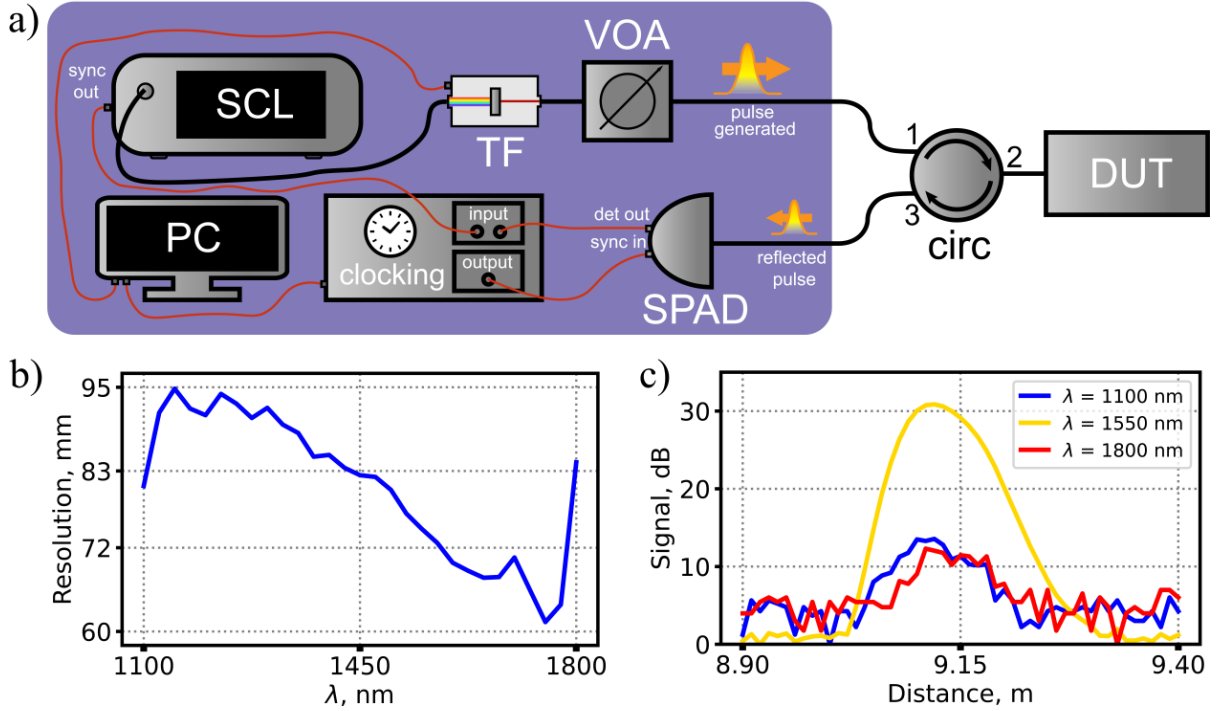

FIG. 2. (a) Broadband $\nu$−OTDR setup. $SCL$, picosecond supercontinuum laser source; $VOA$, variable optical attenuator; $TF$, tunable acousto-optical spectral filter; $CIRC$, optical circulator; $SPAD$, single-photon avalanche photodiode based on InGaAs; $DUT$, device under test (Alice's or Bob's setup). The black lines are optical fibers, and the red lines are electrical wires. (b) The spatial resolution of the broadband $\nu$−OTDR setup across the studied spectral range. (c) The typical reflection signals observed for wavelengths $\lambda = 1100\ nm$ (blue curve), $\lambda = 1550\ nm$ (yellow curve) and $\lambda = 1800\ nm$ (red curve) during OTDR-measurements. These signals are depicted relatively to their noise floor to be compared.

$$R[\text{dB}] = 10 \cdot \log_{10}\left(\frac{P_\text{out}}{P_\text{in}}\right). \quad (6)$$

The $P_{in}$ value can be measured by connecting the VOA output straight to the SPAD. In our setup, the transmittance of the circulator varies with wavelength, which affects the OTDR measurement procedure. Therefore, additional measurements of the circulator spectral transmittance $T_{1\to 2}$ and $T_{2\to 3}$ in the directions $1 \to 2$ and $2 \to 3$ have been conducted using the technique presented previously [19]. Thus, considering the utilization of the same SPAD and SCL for both input ($P_\text{in}$) and output ($P_\text{out}$) measurements, the reflection value $R$ can be estimated as follows:

$$R[\text{dB}] = 10 \cdot \log_{10}\left(\frac{N_\text{out}(l)}{N_\text{in}}\right) + Att_\text{in} \\ - Att_\text{out} - T_{1\to 2} - T_{2\to 3}, \quad (7)$$

wherein $N_\text{out}(l)$ represents the photon counts accumulated during the observation time corresponding to the reflection from a particular region of the DUT located at the distance $l$ from the input, $N_\text{in}$ is the photon counts for the input radiation, $Att_\text{in}$ and $Att_\text{out}$ are attenuation levels installed on VOA to decrease the input and output power. The procedure described in this section is conducted for each wavelength inside the spectral region $\lambda = 1100$ — $1800\ nm$.

## III. EXPERIMENTAL ANALYSIS

### A. Broadband OTDR-traces

We have picked up the optimal operation parameters of the OTDR setup (the laser repetition frequency $f_\text{pulse}$ and the SPAD dead time $\tau_\text{d}$) to set a minimum dark count rate and the measurement duration. Moreover, there are two conditions that such operation parameters should simultaneously satisfy. At one extreme, the SPAD with quantum efficiency $\eta$ must satisfy the condition that it mainly absorbs no more than one photon per detection event, otherwise the count rate $N$ is not linearly dependent on the number of reflected photons:

$$N \leq \eta \cdot f_\text{pulse} \quad (8)$$

For given $f_\text{pulse}$ one can satisfy condition (8) by properly setting the attenuation levels on VOA,

considered in eq. (7). At another extreme, during the dead time set on SPAD no photons must be absorbed, otherwise they will not be detected:

$$\tau_\mathrm{d} \leq \frac{1}{f_\mathrm{pulse}}. \qquad (9)$$

Therefore, constantly monitoring the satisfaction of the conditions (8) and (9), we can vary $f_\mathrm{pulse}$ and $\tau_\mathrm{d}$ to achieve a tradeoff between measurement duration and dark count rate. Optimal operation parameters were found to be $f_\mathrm{pulse} = 500$ kHz, $\tau_\mathrm{d} = 2$ μs, which lead to the dark count rate $N_\mathrm{dark} \approx 1700$ cps and upper photon count limit $N = 50000$ cps.

The QKD optical setups under test are shown in FIG. 3. We deliberately tested the setups without their passive optical defense components to sustain the dynamic range. Spectral transmission of such devices has been previously reported [19], which makes it possible to further account for their operation across a wide spectral range.

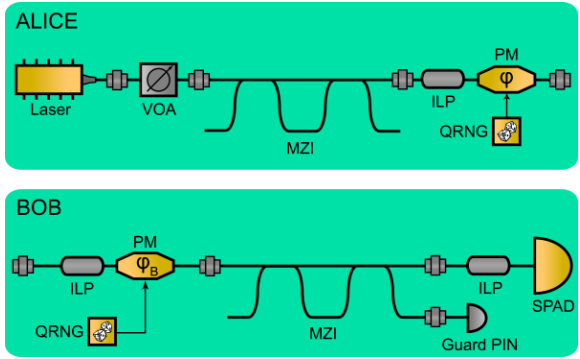

FIG. 3. Alice's and Bob's optical setups under test. VOA, variable optical attenuator; MZI, Mach-Zehnder interferometer; ILP, in-line polarizer; PM, phase modulator; QRNG, quantum random number generator; SPAD, single-photon avalanche photodiode; Guard PIN, pin-diode; and unmarked metal couplers are optical adapters with coupled connectors.

The measurements results for Alice's and Bob's setups are presented in FIG. 4 as a broadband OTDR heatmap across wavelength range $\lambda = 1100 - 1800$ nm. As seen, almost every optical component implemented in the setups under test reflects a non-zero amount of radiation. The maximum reflection value equals $R_{max} \approx -50$ dB at the 9 m distance for Alice's setup and 11 m for Bob's setup explicitly occurs after the Mach-Zender interferometer, which can be detected by the intrinsic pattern of triple peaks due to four light propagation paths with two of them equal.

The correspondence between the reflection peaks and the optical components is presented in the OTDR pictures at the separate wavelength $\lambda = 1325$ nm (see FIG. 5), in which all the reflection peaks are represented clearly. As seen, the maximum pulse reflections after the PM are caused by the fiber connectors. It is reasonable to question why the reflection values obtained from the same components, i.e., the fiber connectors, are significantly different. The fiber connector adapters that couple two fiber endfaces in fact represent a mechanical joint and thereby have certain drawbacks, such as different return losses for different samples. These drawbacks are minor for conventional fiber-optic communications, but they can affect the robustness of the QKD operation in the presence of THA.

Owing to the OTDR device presented in this paper we have obtained spectral reflection values in the time domain from a real QKD system for the first time. The obtained results allow us to further estimate the upper bound for information available to an eavesdropper launching the broadband THA.

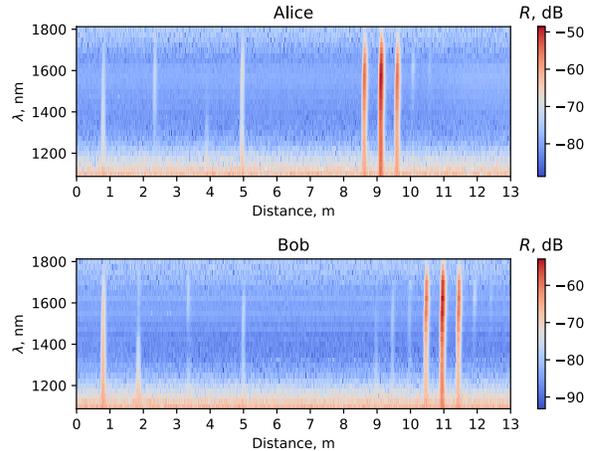

FIG. 4. Broadband OTDR results for Alice's and Bob's optical setup. $\lambda$, the wavelength. The distance along the setup under test is calculated by time delay between the injection of laser pulse inside the setup and the pulse detection.

### B. Optical FC/PC connector analysis

Our QKD setup under test implemented the typical physical contact (PC) connectors. As the OTDR of the Alice and Bob optical setup has shown, the most significant return losses are caused by the optical connectors. Therefore, we decided to conduct OTDR measurements for the separate fiber connector extracted from the QKD system with SPAD running in gated mode. The results of the OTDR measurements for the PC-type optical connector (FIG. 6) show that the reflection value gradually decreases for approximately 9 dB across the wavelength range studied. The noise floor, i.e. Rayleigh-backscattered radiation within the fiber piece corresponding to the gate width, combined with dark counts, was also measured.

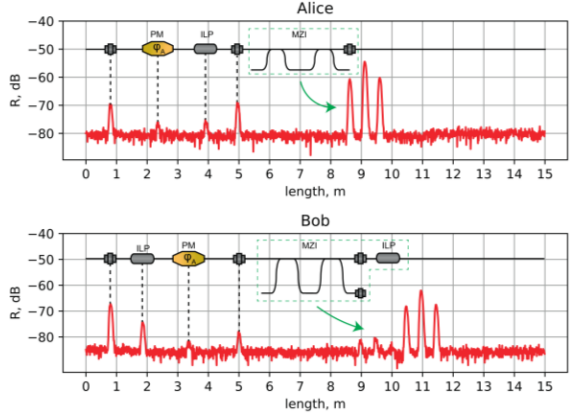

FIG. 5. OTDR results at the wavelength $\lambda = 1325\ nm$ for Alice's and Bob's optical setup and the reflection peaks interpretation. Such devices as SPAD, Alice's laser, VOA and guard PIN are not depicted due to the reflection absence detected.

This reflection value dependence on wavelength can be explained by the internal structure of the PC-type connector. Despite the strong screw joints that are designed to eliminate the air-filled gap between fiber endfaces, small gap may occur due to certain contaminations and microdefects of the fiber core, which are usually removed during the manufacturing stage by quality polishing. However, there is an intrinsic damaged layer with high refractive index, caused by high pressure on the fiber endface during the polishing process [26]. Therefore, the connector behaves like a Fabry-Perot fiber cavity [27]-[28] and the reflection coefficient $R$ (10) – (11) can be determined from the Airy distribution [29]:

$$R_0 = \left(\frac{n_{core} - n_d}{n_{core} + n_d}\right)^2, \quad (10)$$

$$R = 10 \cdot \log_{10}\left(2R_0 \cdot \left(1 - \cos\left(\frac{4\pi \cdot n_d \cdot 2h}{\lambda}\right)\right)\right) \approx$$
$$10 \cdot \log_{10}\left(R_0 \cdot \left(\frac{4\pi \cdot n_d \cdot 2h}{\lambda}\right)^2\right) \quad (11)$$

wherein $n_d$ and $n_{core} = 1.454$ are refractive indices of damaged layer and core, $h$ is the thickness of the damaged layer, $\lambda$ is the wavelength. As previously reported [28], the typical values of the damaged layer refraction index and thickness are inside of ranges $1.46 < n_d < 1.6\ ;0 \leq h \leq 0.11\ \mu m$. Therefore, we can consider $h \ll \lambda$ across the measurement spectral range, which is accounted in (11).

The theoretical model based on (10) − (11) allows to assess the reflection values for the PC-type connector across the near-infrared range. The typical theoretical spectral reflection curve with appropriate parameters $h = 0.015\ \mu m, n_d = 1.474$ is presented in FIG. 6. Additional loss factors originated from imperfect physical contact i.e. misalignments and tilts of the fiber endfaces [27] are determined by the conditions of the experiment, such as a comprehensive force and the quality of the endface polishing. Thus, they are hard-to-control factors and were not considered here. As seen (FIG. 6), the measured reflection values from the PC-type connector across the $1100 − 1800$ nm range seem plausible because they are of the same order and spectral dependence character as predicted by the expressions (10) − (11) based on Fabry-Perot cavity model. The results obtained make it possible to further consider the distinctions of light reflection from fiber connectors in the design stage of the QKD systems to mitigate the eavesdropping by the THA channel.

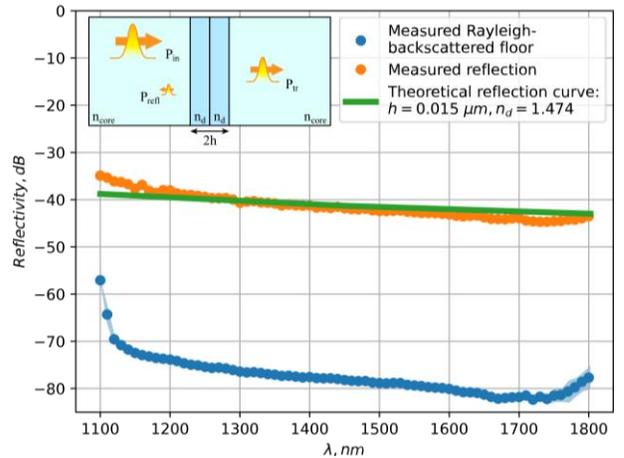

FIG. 6. The reflection values measured by OTDR-device and the calculated reflection values using (10 − 11). $\lambda$ is the wavelength. The upper-left picture shows fiber connector as a physical contact of two fiber endfaces with a damaged layer of the refractive index $n_d$ and the thickness $h$.

We stress that the experimental results obtained in this section drastically depend on initial experimental conditions. The only restriction of the manufacturing standards for FC/PC fiber connectors is that their return losses must be no more than -40 dB at wavelength $\lambda = 1550$ nm. Thus, the specific fiber adapter and connector sample may additionally influence the measured return losses. For instance, in Ref. [31] the authors have shown how the value of the compressive force between fiber endfaces affects the return loss from fiber connector due to the elasto-optical effect. This effect is a hard-to-control factor and thus emphasizes the importance of reflectometry procedures applied to QKD apparatus such as fiber connectors.

The established way to reduce the back-reflections from fiber joints is a high-quality fusion of fiber endfaces, which complicates the reassembling of the optical setup. Nevertheless, the insertion losses sufficiently less than 0.01 dB and return loss less than -80 dB [32] may be achieved by such quality fusion joints. We have observed that return losses of one fusion joint sample lie out of our dynamic range, i.e. -80 dB at single wavelength $\lambda = 1550$ nm. Oblique fiber endfaces also contribute to reducing return losses but rely on the reflected beam withdrawal outside of the fiber core. Theoretically, this withdrawn radiation can be gathered by the eavesdropper. A way to improve the PC-type connector is applying solid resin as refractive index matching material, eliminating the air-filled gap [30].

## IV. THEORETICAL SECURITY ANALYSIS OF THA

The obtained OTDR traces enable us to estimate the information leakage to the eavesdropper. As discussed in Sec. II, the upper bound of the information leakage to the eavesdropper depends on the square roots of the fidelities of Eve's states that originated from the launching of the THA and collective attack $(1 - 3)$. Considering the zero-QBER $Q = 0$ case here we will focus on the leakage value contributed solely by the THA side channel. The final expression for the upper bound of the Holevo value, that is, the mutual information between the legitimate user and Eve, in case of small $\mu_{\text{refl}}$ is as follows [19]:

$$\bar{\chi}_{\text{Eve}} = h\left(\frac{1-\eta}{2}\right) \approx h(\mu_{\text{refl}}) \qquad (12)$$

As discussed in Sec. III, the maximum reflection value $R^A_{\max} \approx -49$ dB occurred at a wavelength $\lambda_A = 1575$ nm in Alice's setup and $R^B_{\max} \approx -53$ dB at $\lambda_B = 1625$ nm in Bob's setup. Considering the absence of watchdog detectors in the optical setup of Alice, the upper bound of Eve's input optical power corresponds to a laser-induced damage threshold equal to approximately 40 dBm [14]. In turn, the optical setup of Bob is equipped by watchdog detector with $-60$ dBm sensitivity, thus non-legitimate radiation power threshold comes down to $-60$ dBm if the THA is launched on the setup of Bob.

The spectral transmittance of the fiber-optical defense components represented a behind-the-scenes aspect of this article, which is already accessed in detail [10], [15]-[19]. For instance, from Ref. [19] we can take a maximum $T_{def}$ value as a sum of the forward and reverse isolator, circulator, and wavelength-division multiplexer (WDM) with the fixed attenuator (only for Alice's setup) transmittance at wavelengths $\lambda_A$ and $\lambda_B$ equaled approximately $-250$ dB for Alice and $-160$ dB for Bob. In this case, the upper bound of a mean photon number value returned to Eve equals $\mu^A_{\text{refl}} \approx 2 \cdot 10^{-16}$ and $\mu^B_{\text{refl}} \approx 8 \cdot 10^{-18}$, which allows us to employ the approximation (12). Therefore, the mutual information between Eve and Alice can be bounded by $\bar{\chi}_{\text{Eve}} \approx 10^{-14}$, and the same for attacking the setup of Bob is $\bar{\chi}_{\text{Eve}} \approx 10^{-16}$, which are apparently negligible information gains for Eve.

## V. CONCLUSIONS

Until today the experimental broadband Trojan-horse attack security analysis has been limited by the spectral transmittance measurement procedure due to the absence of commercially available broadband reflectometry setups. The only way to conduct such an analysis was by making a non-physical assumption of the total broadband reflection of Eve's photons. Such an assumption sometimes results in the necessity of building additional passive optical components in the optical setups of legitimate users to provide the required isolation bound [25]. The incompleteness of such an approach has been a stumbling block in the experimental security analysis of the THA.

In this paper, we have presented an OTDR-measurements technique based on supercontinuum laser source and single-photon avalanche photodiode for experimental estimation of spectral reflectance inside the QKD setups. Our setup enabled us to obtain the broadband OTDR traces that show the largest reflections of approximately -50 dB occurred from the optical connector near the legitimate wavelength $\lambda = 1550$ nm. Eve's information gain turned out to be significantly less than 1, indicating the sustainability of the QKD setup under test to the broadband THA. In principle, the obtained spectral reflection values make it possible to reduce the total QKD setup isolation level to 50 dB.

We stress that complete experimental THA security analysis is only possible by conducting spectral transmittance and reflectance measurements in tandem. Further improvement of the mentioned procedures may be related to the extension of the spectral range to the ultraviolet, visible, and middle-infrared regions, which requires non-conventional light sources and detectors for fiber-based QKD. Furthermore, the measurements under the cutoff-wavelength range with the non-single mode radiation propagating in optical fiber add to the challenge.

## ACKNOWLEDGMENTS
K.D.B. and I.S.S. thank S.P. Kulik, A.N. Klimov and S.N. Molotkov for academic support and fruitful discussions. I.S.S. also acknowledges the scholarship from the Foundation for the Advancement of Theoretical Physics and Mathematics "BASIS".

## APPENDIX

In Ref. [19] an approach to estimate the lower bounds of Uhlmann fidelities between THA states assumed a conservative estimation. It was shown that for mixed phase-coded states such fidelity bounds seem to be unreachable. Here, we will derive an explicitly reachable fidelity bounds for the arbitrary states.

Let us consider THA phase-coded states corresponding to different bits, same as in Ref. [19]:

$$\rho^0 = \sum_{m,n=0}^{\infty} \rho_{mn} |m\rangle\langle n| \qquad (A1)$$

$$\rho^1 = \sum_{m,n=0}^{\infty} \rho_{mn} e^{i(\varphi_m - \varphi_n)} |m\rangle\langle n| = \sum_{m,n=0}^{\infty} \rho_{mn} e^{i\Delta\varphi_{mn}} |m\rangle\langle n|, \qquad (A2)$$

where $|m\rangle$ represents a Fock state with $m$ photons, $\varphi_m$ is a phase shift for a corresponding $m$-photon state.

The focal point of the attention is a lower bound of the square root of the Uhlmann fidelity $\eta = tr\left(\sqrt{\sqrt{\rho^1}\rho^0\sqrt{\rho^1}}\right)$ between such states. Consider $\sigma^0$ and $\sigma^1$ to be the square roots of $\rho^0$ and $\rho^1$:

$$\rho^0 = \sigma^0 \cdot \sigma^0 = \sum_{m,n=0}^{\infty} \left(\sum_k \sigma_{mk}^0 \cdot \sigma_{kn}^0\right) \cdot |m\rangle\langle n|, \qquad (A3)$$

$$\rho^1 = \sigma^1 \cdot \sigma^1 = \sum_{m,n=0}^{\infty} \left(\sum_k \sigma_{mk}^1 \cdot \sigma_{kn}^1\right) \cdot |m\rangle\langle n| = \sum_{m,n=0}^{\infty} \left(\sum_k \sigma_{mk}^0 \cdot \sigma_{kn}^0\right) \cdot e^{i\Delta\varphi_{mn}} \cdot |m\rangle\langle n|. \qquad (A4)$$

Then we express $\Delta\varphi_{mn}$ as $\varphi_m - \varphi_k + \varphi_k - \varphi_n = \Delta\varphi_{mk} + \Delta\varphi_{kn}$, resulting in:

$$\rho_1 = \sum_{m,n=0}^{\infty} \left(\sum_k \sigma_{mk}^0 e^{i\Delta\varphi_{mk}} \cdot \sigma_{kn}^0 e^{i\Delta\varphi_{kn}}\right) \cdot |m\rangle\langle n|. \qquad (A5)$$

Comparing (A5) and (A4) we find the relation between $\sigma^0$ and $\sigma^1$:

$$\sigma_{mn}^1 = \sigma_{mn}^0 \cdot e^{i\Delta\varphi_{mn}}. \qquad (A6)$$

Returning to the fidelity square root expression and using Hermiticity of the density matrices square roots, we obtain:

$$\eta = tr\left(\sqrt{\sqrt{\rho^1} \cdot \rho^0 \cdot \sqrt{\rho^1}}\right) = tr\left(\sqrt{\sigma^1\sigma^0\sigma^0\sigma^1}\right) = tr\left(\sqrt{\sigma^1\sigma^0 \cdot (\sigma^1\sigma^0)^\dagger}\right) = tr(\sigma^1\sigma^0) = tr\left(\sum_{m,n=0}^{\infty} \left(\sum_k \sigma_{mk}^0 \sigma_{kn}^0 \cdot e^{i\Delta\varphi_{mk}}\right) |m\rangle\langle n|\right) = \sum_{m,k=0}^{\infty} \sigma_{mk}^0 \sigma_{km}^0 \cdot e^{i\Delta\varphi_{mk}}, \qquad (A7)$$

where † is Hermitian conjugation. Note that the first term in the final sum corresponds to a vacuum component probability $p_0 = (\sigma_{00}^0)^2$, thus the final expression for the fidelity square root bound is:

$$\eta = p_0 + \sum_{m,k=1}^{\infty} \sigma_{mk}^0 \sigma_{km}^0 \cdot e^{i\Delta\varphi_{mk}} \geq p_0 - \sum_{m,k=1}^{\infty} \sigma_{mk}^0 \sigma_{km}^0 = p_0 - (1 - p_0) = 2p_0 - 1 \geq 1 - 2\mu_{Eve} \qquad (A8)$$

where we set $\Delta\varphi_{mk} = \pi$ for every $m, n$ indices values for lower bounding. This bound coincides with the pure states bound derived in Ref. [19]. The optimal THA states are as follows:

$$\xi^0 = \sqrt{1 - \mu_{Eve}} \cdot |0\rangle + \sqrt{\mu_{Eve}} \cdot |1\rangle,$$
$$\xi^1 = \sqrt{1 - \mu_{Eve}} \cdot |0\rangle - \sqrt{\mu_{Eve}} \cdot |1\rangle.$$

This result explicitly indicates that there is no benefit for Eve in launching the THA using mixed states. Nevertheless, the refined fidelity lower bound does not detract from the results obtained in Ref. [19], since both bounds are asymptotically equal, when $\mu_{Eve} \ll 1$, which is often the case.